\documentclass[sigconf,balance=false]{acmart}
\usepackage{popets}

\usepackage{graphicx}
\usepackage{xcolor}
\usepackage[font=small]{caption}
\usepackage{tabularx}
\usepackage{multirow}
\usepackage{enumitem}
\usepackage{pifont}
\newcommand{\xmark}{\text{\ding{55}}}
\newcommand{\cmark}{\text{\ding{51}}}
\newcommand{\myparatight}[1]{\vspace{0.5ex}\noindent\textbf{#1.}}
\usepackage{float}
\restylefloat{table}
\usepackage{subfig}

\setcopyright{popets}
\copyrightyear{YYYY}

\acmYear{YYYY}
\acmVolume{YYYY}
\acmNumber{X}
\acmDOI{XXXXXXX.XXXXXXX}
\acmISBN{}
\acmConference{Proceedings on Privacy Enhancing Technologies}
\begin{document}

\title{SoK: Secure Human-centered Wireless Sensing}

\author{Wei Sun}
\orcid{1234-5678-9012}
\affiliation{%
  \institution{Duke University}
  \city{} 
  \state{} 
  \country{} 
}
\email{redsunwit@gmail.com}

\author{Tingjun Chen}
\affiliation{%
  \institution{Duke University}
  \city{}
  \country{}}
\email{tingjun.chen@duke.edu}

\author{Neil Gong }
\affiliation{%
  \institution{Duke University}
  \city{}
  \country{}
}
\email{zhenqiang.gong@duke.edu}

\renewcommand{\shortauthors}{Sun et al.}

\begin{abstract}
Human-centered wireless sensing (HCWS) aims to understand the fine-grained environment and activities of a human using the diverse wireless signals around him/her. While the sensed information about a human can be used for many good purposes such as enhancing life quality, an adversary can also abuse it to steal private information about the human (e.g., location and person's identity). However, the literature lacks a systematic understanding of the privacy vulnerabilities of wireless sensing and the defenses against them, resulting in the privacy-compromising HCWS design.

In this work, we aim to bridge this gap to achieve the vision of secure human-centered wireless sensing. First, we propose a signal processing pipeline to identify private information leakage and further understand the benefits and tradeoffs of wireless sensing-based inference attacks and defenses. Based on this framework, we present the taxonomy of existing inference attacks and defenses. As a result, we can identify the open challenges and gaps in achieving privacy-preserving human-centered wireless sensing in the era of machine learning and further propose directions for future research in this field.
\end{abstract}


\maketitle

\section{Introduction}
\label{sec:intro}

Wireless sensing is an emerging enabling technology for many applications such as smart homes/cities, autonomous systems, and human-computer interactions. Given the advanced wireless communication techniques (\emph{e.g.}, WiFi, and 5G) and the proliferation of wireless devices (\emph{e.g.}, Internet-of-Things), wireless sensing is becoming more and more popular. Wireless signals in different forms, including \emph{radio frequency (RF)} and \emph{light}, interact with human bodies and other physical objects in the environment during transmission. As a result, the variation of the wireless signals around a human can be leveraged to understand the physical environment and human activities in it~\cite{adib2013see, wang2014rf, vasisht2018duet, zhu2020tu}. For instance, Vasisht et al.~\cite{vasisht2018duet} shows that wireless signals can be used to localize and identify occupants at home based on their walking patterns, thereby enabling a smart home that is aware of the occupants' locations and identities to personalize appliance settings.

Like nearly any advanced technology, wireless sensing is a double-edged sword. On the one hand, wireless sensing enables many life-quality-improving applications such as health status monitoring~\cite{gu2018sleepy, ha2021wistress, abedi2020wifi}, energy-efficient smart home~\cite{pu2013whole, vasisht2018duet, cohn2012humantenna}, and friendly human-computer interaction~\cite{wang2014rf, luo2021rfaceid} via understanding the physical environment and activities of human subjects. On the other hand, the same technology can be abused by an attacker to infer a human's private information such as location, living habits, and behavioral biometric characteristics (\emph{e.g.}, walking pattern, heart rate, and hand gesture) that can identify a person, therefore leading to privacy and security risks. For instance, inferring location leads to location privacy leakage~\cite{vasisht2018duet, wang2013dude}; inferring living habits may lead to well-planned burglary~\cite{zhu2020tu, staat2022irshield}; and inferring hand gesture used to unlock a smartphone leads to password compromise~\cite{ali2015keystroke,li2016csi}.

However, the literature lacks a systematic understanding of inference attacks via wireless sensing and defenses against them. In particular, existing literature surveys about wireless sensing \cite{liu2019wireless, mamdouh2018securing, wang2019physical, shrestha2022sok} focus on wireless sensing techniques and their \emph{benign} applications, leaving systematization of the privacy aspect of wireless sensing largely untouched. Such a gap makes it hard to comprehensively understand the privacy vulnerabilities of wireless sensing and design effective defenses against potential inference attacks in the future. Without comprehensive systematization of wireless sensing systems, it is difficult for engineers to design privacy-preserving wireless sensing systems.

In this paper, we aim to bridge this gap. To do so, we propose a signal processing pipeline to systematize the inference attacks and defenses in human-centered wireless sensing systems. More specifically, we make the following contributions:
\begin{itemize}
    \item \textbf{Taxonomy of wireless signals processing in the inference attacks and defenses.} Since wireless signal processing has been extensively used in human-centered wireless sensing systems for inference attacks and defenses, we propose a generalized signal processing pipeline-based framework for reasoning the existing and future inference attacks and defenses. 
    \item \textbf{Open challenges.} We use our proposed framework to identify significant challenges facing the existing human-centered wireless sensing systems, predict the potential inference attacks, and provide directions for potential defenses against these attacks.
    \item \textbf{Identifying the design space towards privacy-preserving wireless sensing.} We identify the core design aspects that future wireless sensing systems should consider in their design to achieve privacy-preserving properties, and provide a design roadmap by discussing where and how human private information has been leaked based on our proposed framework.
\end{itemize}

\begin{table*}
\resizebox{\textwidth}{!}{\begin{tabular}{|c|c|c|c|}
\hline
\textbf{}                     & \textit{\textbf{Mobile computing community}}                                                                                                         & \textbf{Security and privacy community}                                                                   & \textbf{others}                                                             \\ \hline
\textbf{Conferences/Journals} & \textit{\begin{tabular}[c]{@{}c@{}}Sigcomm, Mobicom, NSDI, Mobisys\\ Sensys, IMWUT/Ubicomp, ToN, \\ TOSNHotnet, IPSN, CoNext, Buildsys\end{tabular}} & \begin{tabular}[c]{@{}c@{}}Security and Privacy, NDSS, CCS\\ WiSec, USENIX Security, AsiaCCS\end{tabular} & \begin{tabular}[c]{@{}c@{}}INFOCOM, TOG, VTC,CHI, ICOIN, ICC, TOBD, ECCV, CVPR, FTTC, IJDSN\\ ToGRS, WiSPNET, Sensors, COMPSAC, TMTT, ICUW, ICASSP, Percom, RadarCon\\ ICCA, communication letters,IEEE surveys and tutorials, Geriatric Psychiatry, RFID-TA \\ JSAC, information systems, TOCS,JSTSP, JFI, GRSL, DySPAN, TOMC, MCM, IoTJ,\\ AJGP, ToIM, WF-IoT, NaNA\end{tabular} \\ \hline
\end{tabular}}
\caption{Summary of surveyed venues.}
\label{appendix:conference:table}
\end{table*}

\section{Method}
\label{sec:method}

To systematize the knowledge of inference attacks and defenses to human-centered wireless sensing, we adopt the five-step iterative process proposed by Wolfswinkel et al.~\cite{wolfswinkel2013using} for literature review, which includes: (1) \emph{Define}, (2) \emph{Search}, (3) \emph{Select}, (4) \emph{Analyze}, and (5) \emph{Present}. 

\myparatight{Define}  We define the scope of our literature review as follows:
\begin{itemize}[leftmargin=*,itemsep=3pt]
    \item
    \emph{Selected Source.} We use Google Scholar, ACM digital library, and IEEE Xplore as sources to collect papers. Moreover, we present the papers based on our research activities and common knowledge from the ACM, IEEE, USENIX, and ISOC communities. We report the papers published in the wireless sensing venues (e.g., SIGCOMM, MobiCom, IMWUT, Mobisys, HotNets, and Sensys) and network security and privacy venues (e.g.,  USENIX Security Symposium, IEEE S and P, and NDSS). Table~\ref{appendix:conference:table} shows the summary of all the surveyed venues.
    \item
    \emph{Search Terms.} We search the papers using the following terms: wireless sensing/localization, human activity recognition, inference/privacy attacks and defenses, eavesdropping, and human-centered wireless sensing. 
    \item
    \emph{Inclusion Criteria.} We mainly include papers from peer-reviewed journal articles as we presented in the Selected Source, which focus on how to \textit{infer human private information in human-centered wireless sensing.} Specifically, we read the paper to understand if the paper's theme matches the human-centered wireless sensing topic. We find that some workshop or arXiv papers have the corresponding full papers published in the official conferences. So, we will simply select the full papers published in the official conferences. Moreover, we will eliminate the papers that do not discuss the \textit{physical-layer wireless sensing techniques for inference attacks or defenses}, as human-centered wireless sensing mainly exploits the interaction between the wireless signals and the human body.  
\end{itemize}

\myparatight{Search} Except for the well-known papers in this area based on our experience, we use the sources (\emph{e.g.}, ACM digital library) mentioned above to search for papers. Moreover, we investigate the references in the related work presented in these papers to further build on the collected knowledge. As a result, we have 184 papers as candidates for the analysis.  

\myparatight{Select} We aim to select the papers in the search stage, which can satisfy the inclusion criteria. Specifically, we first read the abstract and introduction sections of each paper to obtain a high-level view of its main discussion and focal points. Then, we read the core design of the paper to ensure the inclusion criteria. At last, there are 169 papers left for our systematization. The other 15 papers cannot meet our inclusion criteria. For example, some papers discuss the differential data privacy of wireless communication traffic.

\myparatight{Analyze} After selecting the papers, we divide them into two categories based on their topics. The first category mainly focuses on designing wireless sensing-based inference attacks that can accurately infer a victim's various private information. The second category mainly focuses on defenses against such attacks.

\myparatight{Present} We present our findings of formalizing the inference attacks and defenses to human-centered wireless sensing as follows.

\section{Threat Model}
\label{sec:threat:model}
\subsection{Attacker's Goal}
We consider an attacker’s goal to infer various private information
about a victim human through sensing and analyzing the
wireless signals around him/her. In particular, we summarize the
private information considered in existing inference attacks as the
following three categories:
\begin{itemize}
    \item \textbf{Location.} The location represents sensitive information about a victim. Knowing the location of a victim leaks sensitive places that the victim has been to, such as those in a hospital, and enables tracking of the victim~\cite{ayyalasomayajula2020deep, chen2021pushing, kotaru2015spotfi}.
    \item \textbf{Living habits.} The living habits of a victim can leak other sensitive information about a victim. For instance, eating meals and going to the restroom frequently could be an indicator of diabetes disease. Moreover, knowing the living habits of a victim enables an attacker to commit well-informed severe crimes. For instance, an attacker may plan a burglary at a time when a victim is not at home~\cite{abedi2020wifi, acar2020peek, adib2015capturing}.
    \item \textbf{Behavioral biometric characteristics.} Behavioral biometric characteristics refer to a person’s pattern of behavior, including walking patterns, heart rate, and hand gestures. The leak of such behavioral biometric characteristics of a victim leads to severe privacy and security risks to the victim. For instance, heart rate may reveal that a victim has asthma or heart disease; hand gesture (e.g., touched locations and swiping patterns on the screen) of a victim to unlock a smartphone leads to compromise of the victim’s password; and walking patterns enable an attacker to identify the victim’s identity~\cite{ali2015keystroke,liu2021hand,jiang2022rf, feng2021rf}.
\end{itemize}

\subsection{Attacker's Capability}

\textbf{Sensing the type of wireless signal.} We consider the attacker can sense the types of wireless signals around a victim. For instance, the attacker can first perform coarse-grained spectrum scanning to check if electromagnetic waves exist in the physical environment and then use fine-grained spectrum scanning to figure out the operating frequency of the wireless signals if they exist.

\noindent \textbf{Receiving wireless signals via deploying a radio receiver.} After the attacker senses the type of wireless signals, we consider the attacker is able to deploy a radio receiver to receive the wireless signals. The radio receiver should not be too far away from the transmitter around the victim, in order to receive wireless signals. For instance, when an attacker targets a victim in a house, the attacker can deploy its radio receiver outside/around the house.

\section{HCWS and its privacy implications}
\label{sec:threat}

\subsection{Wireless Sensing Principle}
\begin{figure}
  \centering
  \includegraphics[width=\linewidth]{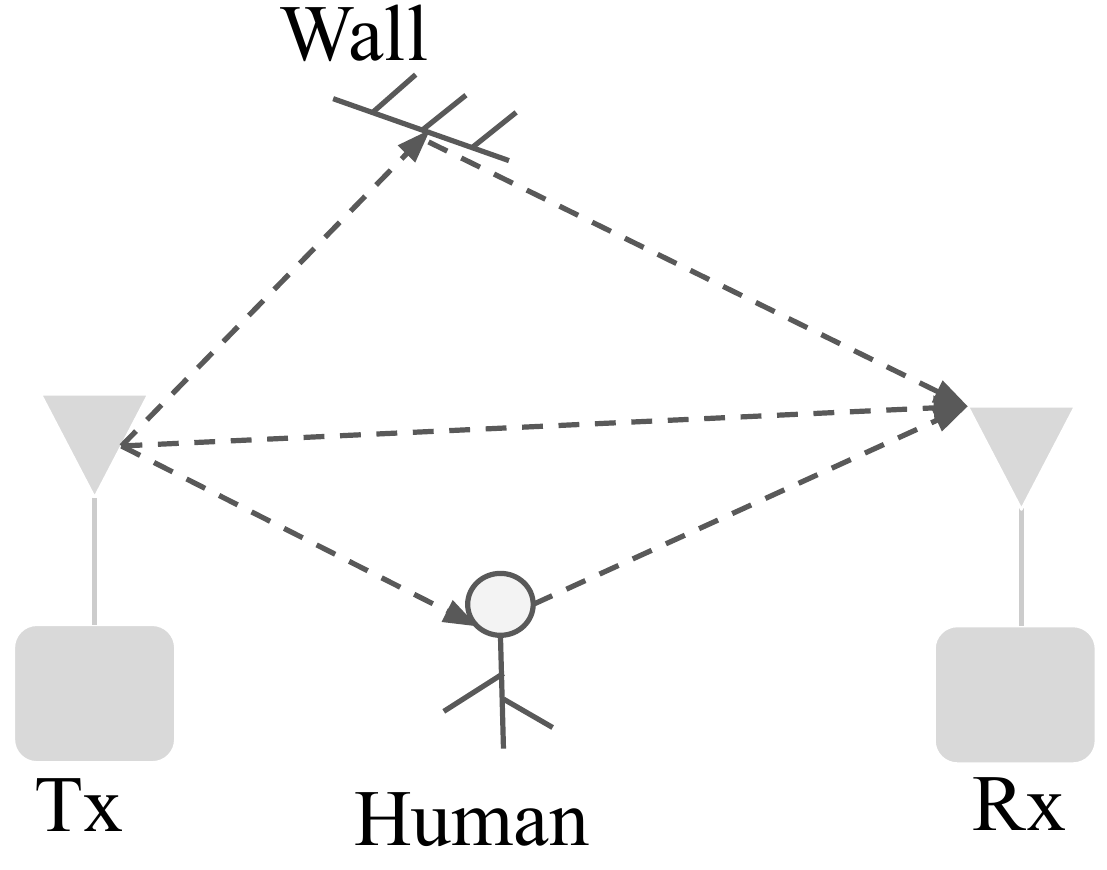} 
    \caption{A typical wireless sensing system consists of a transmitter (Tx) and a receiver (Rx), where the Tx transmits wireless signals undergoing the physical environment and the Rx receives wireless signals. The wireless signals may reach the Rx through multiple paths  due to  reflections of the different objects (e.g., walls) and subjects (e.g., humans) in the physical environment.}
    \label{fig:wireless:sensing}
\end{figure}

A typical wireless sensing system consists of two devices: a \emph{transmitter (Tx)} and a \emph{receiver (Rx)}, as shown in Fig.~\ref{fig:wireless:sensing}. A Tx or Rx may have one or multiple \emph{antennas}. A Tx antenna emits wireless signals, which propagate and may be reflected by different objects (\emph{e.g.}, walls) and subjects (\emph{e.g.}, human) in the physical environment. An Rx antenna receives wireless signals. 

To model the wireless communication between a Tx and an Rx, we start with a pair of Tx and Rx, each equipped with a single antenna. Specifically, the Tx transmits the wireless signals, denoted by $x(t)$, which is reflected by different types of objects (\emph{e.g.}, walls, desks, and couches) and subjects (\emph{e.g.}, human) in the physical environment, and then received by the Rx. Let $h(t)$ denote the multipath propagation characteristics of the physical environment, or the \emph{wireless channel}.

\begin{table}
\resizebox{0.5\textwidth}{!}{\begin{tabular}{|c|c|c|c|}
\hline
\textbf{Wireless technology} & \textit{\textbf{Cost}} & \textbf{Effectiveness} & \textbf{Deployability} \\ \hline
\textbf{WiFi}                & \textit{Medium}        & High                   & High                   \\ \hline
\textbf{BLE/Zigbee}          & Low                    & Medium                 & High                   \\ \hline
\textbf{RFID}                & Low                    & Low                    & High                   \\ \hline
\textbf{mmWave/UWB radar}    & High                   & High                   & High                   \\ \hline
\textbf{VLC}                 & Low                    & Low                    & Low                    \\ \hline
\textbf{Cellular}            & Medium                 & High                   & Low                    \\ \hline
\end{tabular}}
\caption{Comparison of wireless technologies.}
\label{table:wireless:tech}
\end{table}

\subsection{Wireless Technologies} 
\label{sub:sec:wireless:tech}
There are many different kinds of wireless technologies that can be used for interference attacks. Table~\ref{table:wireless:tech} summarizes the cost, effectiveness, and deployability of different wireless technologies. More details about these wireless technologies can be found as follows:
\begin{itemize}
\item \myparatight{WiFi} WiFi has been extensively explored for human-centered wireless sensing by harnessing the existing WiFi communication infrastructure~\cite{guo2017wifi, li2016wifinger, jiao2021openwifi}. To conduct the inference attacks using WiFi signals, the attacker needs to extract the channel state information (CSI) from the network interface card of the commercial WiFi device (e.g., laptop) or software-defined radios (e.g., USRP). Therefore, WiFi-based inference attacks do not introduce extra deployment costs and are easy to conduct with open-sourced CSI extractors~\cite{intel_wifi_card, atheros_wifi_card}. 

\item \myparatight{BLE/Zigbee} Bluetooth low energy or Zigbee is designed for short-range and low-power communication, which can be also leveraged for inference attacks~\cite{givehchian2022evaluating, shrestha2022sok,ayyalasomayajula2018bloc}. As the BLE/Zigbee-enabled devices (e.g., mobile devices) are widely deployed and the BLE/Zigbee sensor is usually low-cost, it is easy to conduct the inference attack. However, BLE/Zigbee suffers from the short communication and sensing range. It requires the attacker to be close to the target of interest. So, the attacker can be easily exposed and defended.  

\item \myparatight{RFID} Passive UHF RFID tags are widely used and deployed in warehouses and grocery stores for internet-of-things applications with a short communication range~\cite{luo2021rfaceid,zhang2010tasa, tanbo2015active, merenda2021device, wang2013dude, piva2021tags}. The UHF RFID tags are low-cost, low-power, and small form factors without instrumenting complicated cryptographic algorithms, which can be used for inference attacks with the RFID reader. We can either use a commercial off-the-shelf RFID reader or software-defined radios to investigate the RFID tags, while the RFID readers are expensive. Since the passive RFID tags can be blindly investigated by any RFID reader, the RFID systems are supposed to be vulnerable to tag ID exposure. 

\item \myparatight{mmWave/UWB Radar} mmWave radar's physical principle is to emit the frequency-modulated continuous waves to the target of interest and analyze the signals reflected back from the target of interest for wireless sensing. Since mmWave is a high-frequency wireless signal, it usually gets attenuated easily~\cite{yu2020mmwave, cho2023mmwall, li2018eye,liu2022mtranssee}. So, the mmWave radar is usually instrumented with a phased array antenna to concentrate the signals within a narrow beam for long-range sensing. The commercial off-the-shelf mmWave radar is usually low-cost and its sensing ability is limited by its phased array antenna. Since mmWave radar usually has a large bandwidth, it can provide very fine-grained sensing accuracy. Ultra-wideband radar (UWB) usually emits an impulse with a large bandwidth and measures the time-of-flight of the signals reflected off the target of interest, which can provide very accurate time-of-flight measurements with a larger bandwidth. In comparison to WiFi, BLE, and RFID, UWB and mmWave sensors are not widely deployed. All these sensors are usually cheap and easy to have from the market. 

\item \myparatight{VLC} Visible light communication usually works at high frequency which is supposed to be significantly attenuated over the air~\cite{cui2020sniffing, li2015human}. Therefore, VLC-based wireless sensing has a short sensing range in comparison to WiFi. However, VLC employs a large bandwidth to measure the time of flight for accurate sensing with ~\cite{li2015human}. To do VLC-based inference attacks, we need to deploy the low-cost LED sensors close to the subject of interest in a line-of-sight scenario and VLC suffers from the interference introduced by the ambient light signals, which makes this VLC-based inference attack impractical in real-world settings.

\item \myparatight{Cellular} Since the cellular communication infrastructure has been widely deployed in outdoor environments, we can use it for inference attacks such as outdoor localization~\cite{kotuliak2022ltrack, vasisht2016eliminating,bae2022watching}. The cellular-based inference attacks in human-centered wireless sensing suffer from the multipath effect in the outdoor area resulting in coarse-grained sensing accuracy. For example, LTrack~\cite{kotuliak2022ltrack} can achieve  6m localization error in 90$\%$ cases. 
\end{itemize}

\subsection{Workflow of Inference Attack}
\label{subsec:workflow}

\textcircled{1} \textbf{Deploying an sensing device.} When the existing wireless sensing system has already been deployed in the environment for good purposes such as enhancing life quality, the attacker can abuse it by deploying a receiver to sniff the wireless signals for human private information inference. Since the wireless signal is transparent to the attacker, the attacker needs to ensure the type of wireless signals used in the environment and choose the corresponding sniffing device to receive the wireless signals. In particular, the attacker can perform spectrum scanning to obtain the type of wireless signals in the environment and their corresponding operating frequency. Spectrum scanning can be divided into two categories: (\emph{i}) using dedicated spectrum analyzers, which have poor time resolution due to large sweeping time~\cite{salous1998digital, spectrum_analyzer}, and (\emph{ii}) using low-cost radio receivers, which have small signal bandwidths due to the limited sampling rate~\cite{hassanieh2014ghz, rashidi2011nlls, shi2015beyond}. Recently, SweepSense~\cite{guddeti2019sweepsense} proposes to modify the software-defined radio receiver (\emph{i.e.}, USRP N210) to sweep the spectrum with high bandwidth and time resolution.

\begin{figure*}
  \centering
  \includegraphics[width=\linewidth]{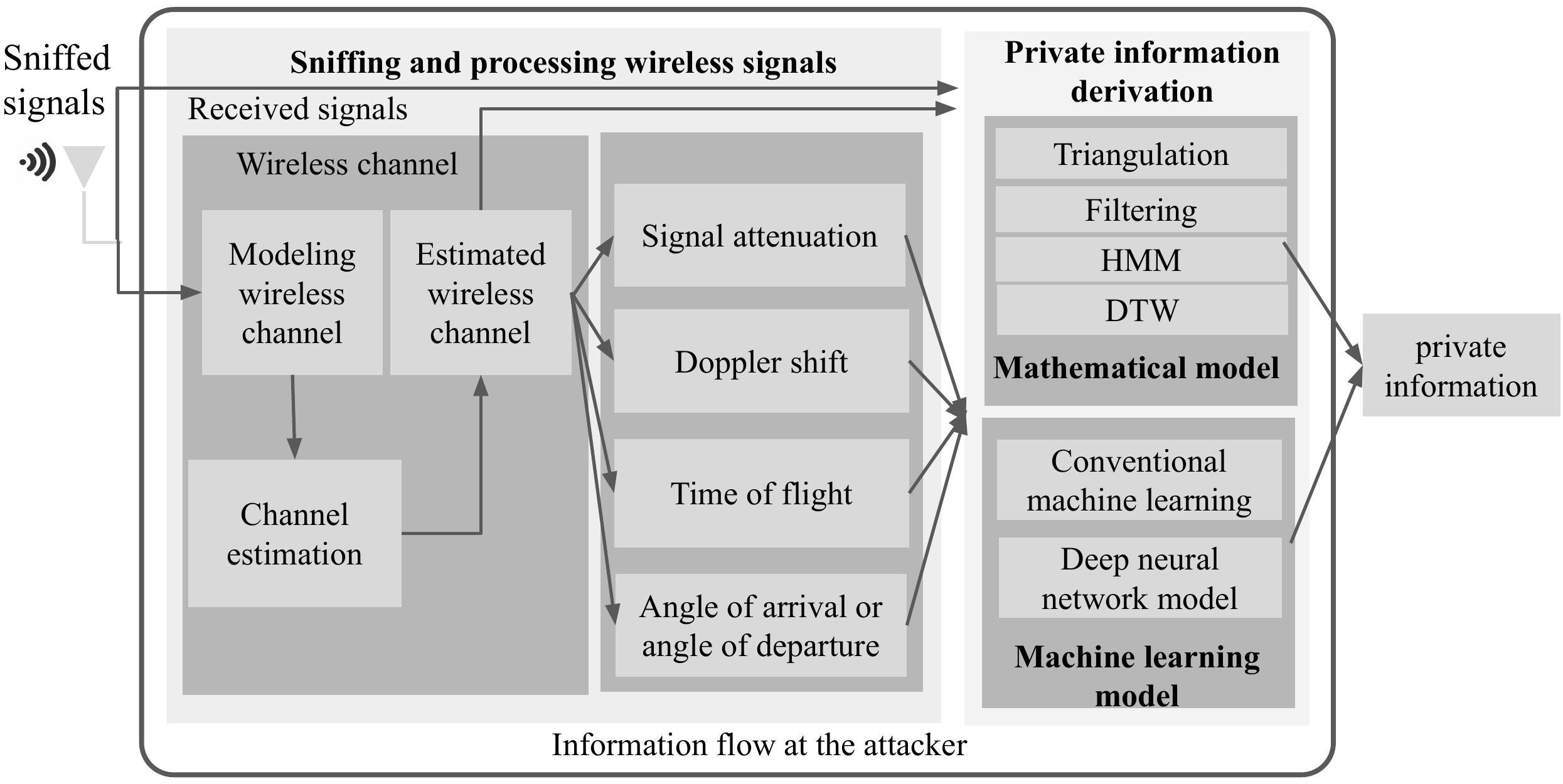} 
    \caption{Overview of our proposed signal processing pipeline-based HCWS framework for analyzing and systematizing the existing human-centered wireless sensing.}
    \label{fig:pipeline}
\end{figure*}
When there are no existing wireless sensing systems deployed in the environment, passive attacks cannot sniff any wireless signals interacting with the human body for inference attacks. However, the active attacker can deploy the transmitter to emit the wireless signals toward the physical environment and receive the backscattered signals to infer human private information, as the emitted wireless signals interact with the human body. To eliminate the multipath effect, we can either leverage the beamforming technique or multipath resolving algorithms. For example, Spotfi~\cite{kotaru2015spotfi} proposes a super-resolution algorithm to estimate the angle-of-arrival (AoA) by incorporating a filtering and estimation approach to accurately identify the AoA of the direct path.

\textcircled{2} \textbf{Sniffing and processing wireless signals.}
The attacker can either actively emit the wireless signals and then receive the backscattered signals or passively receive the ambient wireless signals from the environment to infer human private information. As the received wireless signals are affected by the subject of interest in the physical environment, it is feasible to predict the human private information from these sniffed wireless signals. Then, the attacker needs to extract the wireless signals that are only affected by the subject of interest by resolving the multipath reflections, as the received signals at the attacker are the results of the multipath effect.

\textcircled{3} \textbf{Inferring human private information.}
After obtaining the wireless signals that are only affected by the subject of interest, the attacker can design a model to predict the human private information from the wireless channel measurements through mathematical analysis or learning-based approaches.

\subsection{Privacy Implications}
\label{subsec:privacy:imp}
\textbf{From wireless sensing to privacy inference.} Wireless sensing aims to perceive the physical environment using the received wireless signals around a human. The intuition is that the received signals are affected by the wireless channel, which is affected by the variation of the wireless environment (\emph{e.g.}, human's movements) as the wireless signals interact with the human body. Therefore, a wireless sensing system usually analyzes the variation and extracts different properties (e.g., wireless channel) of the received signals to achieve the sensing purpose, which can reveal human private information such as human location and identity.

\noindent \textbf{Bridging the gap.} However, the existing wireless communication standards and specifications fail to prevent the leakage of this information due to the nature of widespread wireless signals. Wireless sensing systems have been extensively studied in academia, which mainly focuses on improving sensing accuracy without considering privacy leakage. The fundamental reason for privacy leakage is the interaction between the human body and wireless signals. Especially, with the proliferation of deep learning-based wireless sensing systems, even though deep learning has significantly improved the sensing accuracy of wireless sensing, it is vulnerable to adversarial attacks~\cite{carlini2017adversarial, jia2018attriguard, jia2019memguard, zhou2022wiadv}. Therefore, we provide an angle of understanding the vulnerability and privacy threat of machine learning-enabled wireless sensing systems from the whole system design point of view. Human private information is not communication data privacy but rather private information related to human movement that is sensed by the variation of the wireless signals. For example, keystrokes and gestures can reveal passwords. Human activity recognition can reveal the daily living style and human identities. The attacker can abuse the private information introduced by the human movement.  For example, the attacker can detect if the house owner is at home or not for trespassing and theft.  
We bridge the gap between the privacy implications and wireless signal sensing parameters by connecting the physical parameters in the signal processing to the privacy inference that is targeted by the attackers.

\section{A Signal Processing Pipeline-based HCWS Framework}

\label{sec:framework}
We present a signal processing pipeline-based framework to categorize and systematize HCWS strategies as shown in Fig.~\ref{fig:pipeline}. As discussed in Section~\ref{subsec:workflow}, the attacker sniffs the wireless signals propagated in the physical environment using the deployed sensing device to extract different information from the wireless signals for human private information derivation. The wireless signal processing pipeline is modeled and framed to systematize the HCWS.

\subsection{Wireless Channel Estimation}
An attacker can reconstruct a wireless channel from the received wireless signals, which will be used to derive human private information. Let's first model the wireless channel. When a device transmits a signal, this signal is distorted by the wireless environment due to human movements. Specifically, the signal undergoes the attenuation $\alpha(t)$ due to path loss and absorption. Since the signal travels over a distance of $d(t)$, its phase and strength can be changed. In a multiple-antenna wireless sensing system, we can consider the extra distance that the signal travels to/from each antenna in comparison to the reference antenna. This is characterized by the angle of arrival (AoA) $\theta_l(t)$  for $l$-th signal path at the antenna array-enabled Rx and the angle of departure (AoD)  $\varphi_l(t)$ for $l$-th signal path at the antenna array-enabled Tx.

The wireless channel $h(t)$ can be obtained using signal preambles known to both the Tx and Rx and indicates the variation of the wireless environment. Let $p(t)$ denote the preamble signal, the received preamble at the Rx is given by:
\begin{equation}
\label{eq:sensing:model}
    y_p(t)=h(t)*p(t)+w(t).
\end{equation}
With the known $p(t)$ and white Gaussian noise $w(t)$, $h(t)$ can be obtained using the maximum likelihood estimator. Based on the assumption that the signals at the adjacent frequency will undergo the same multipath, ML-based channel estimation methods have also been proposed in~\cite{bakshi2019fast, liu2021fire, vasisht2016eliminating, jiang2018towards}. 

\subsection{Human Private Information Inference}
\label{subsec:hpii}
To infer private information related to the victim, we need to find the relationship between the desired human private information and the extracted features from the received wireless signals. Prior works on human private information derivation mainly focus on the following methods.

\myparatight{Triangulation} The location of the victim can be obtained through triangulation, which can leverage the features from multiple receiving devices deployed by the attacker. Then, the wireless signals' features from these receiving devices deployed by the attacker can be used to reduce the ambiguity due to the noise. For example, the overlap of two features (\emph{e.g.}, AoAs) can pinpoint the location of the victim~\cite{wang2014rf}. The feature (\emph{e.g.}, ToF) from one receiving device deployed by the attacker can formulate an ellipse. The overlap of multiple ellipses can pinpoint the location of the victim~\cite{adib20143d, adib2015multi, luo20193d}.

\myparatight{Filtering}
To obtain the location of the victim, the attacker can use filters to filter out the extracted features that are not related to the victim. The widely used filtering methods for localization, tracking, and gesture/activity recognition include Kalman filtering and particle filtering. For example, TurboTrack~\cite{luo20193d} leverages particle filtering to achieve robot localization. Pantomime~\cite{shangguan2017enabling} uses extended Kalman filtering to achieve gesture recognition.

\myparatight{Markov chain modeling}
Since tracking, hand gestures and human activity recognition are time-series movements, it is intuitive to leverage Markov chain models to delineate these time-series events. Prior works mainly use the Markov chain model or hidden Markov model (HMM) for tracking, localization, and gesture recognition. For example, TurboTrack~\cite{luo20193d} uses HMM to track RFID-tagged drones. Lei et al.~\cite{yang2015see} use HMM to track moving objects through the wall.

\myparatight{Dynamic time warping (DTW)}
The main idea of DTW is to measure the similarity between the extracted and ground-truth features for human private information inference. For example, Mudra~\cite{zhang2016mudra} uses DTW to recognize hand gestures, and Holt et al.~\cite{ten2007multi} leverage the multi-dimensional DTW for hand gesture recognition.

\myparatight{Machine learning models}
The machine learning model, especially the deep neural network,  has been widely used to infer human private information due to its powerful data representation, resulting in highly accurate human private information derivation. Therefore, recent works on human-centered wireless sensing mainly design deep neural networks for highly accurate human private information derivation~\cite{ayyalasomayajula2020deep,ha2021wistress,koutris2022deep}. However, these machine-learning models are suffering from cyber attacks, the large training dataset collection, and scalability. Especially, in the wireless sensing domain, as the wireless environment is dynamic and full of multipath, it is very challenging to have well-trained and trustworthy machine learning models for human-centered wireless sensing~\cite{liu2023exploring}.

\begin{table*}
  \centering
  \includegraphics[width=0.8\linewidth]{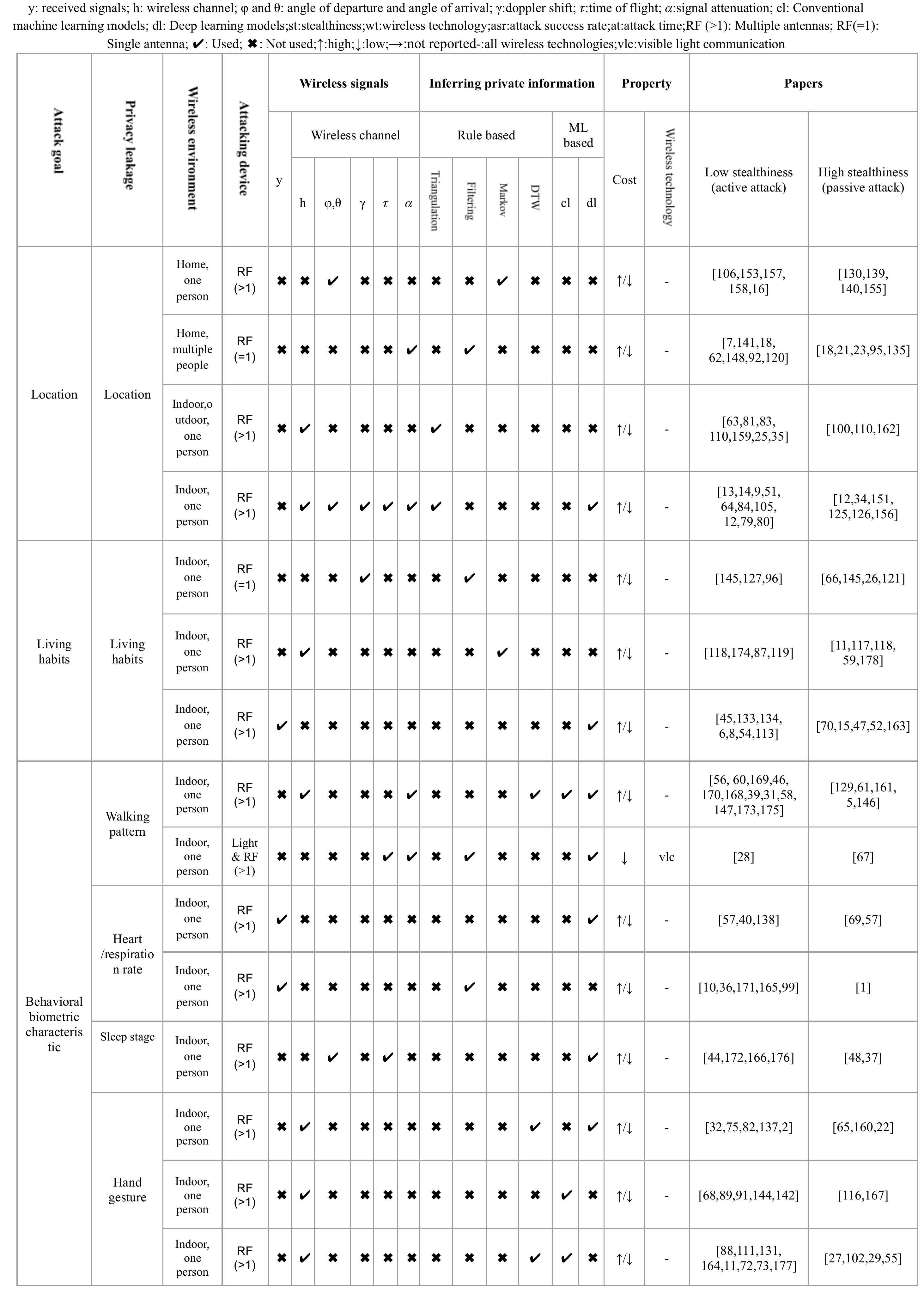} 
   \caption{Taxonomy of existing inference attacks in the human-centered wireless sensing.}
    \label{fig:sys:attack}
\end{table*}

\section{Taxonomy of Existing Inference Attacks}
\label{sec:attacks}

Table~\ref{fig:sys:attack} shows the taxonomy of existing inference attacks based on our proposed signal processing pipeline-based framework, where the sniffed wireless signals at the attacker will be processed and distilled to infer human private information. The prior works are categorized across multiple dimensions such as the attack goal, privacy leakage, wireless environment, attacking device, wireless signals, inferring private information, and property. Note that the property includes three metrics: cost, stealthiness, and wireless technology. The cost metric is measured by whether the attack requires a customized hardware device that can work with high bandwidth or a large antenna array. Usually, the customized attacking device working at the high bandwidth with a large antenna array is high-cost. The ubiquitous wireless radios such as WiFi access points and COTS software-defined radios are considered to be low-cost and are widely available.  The wireless technology column indicates all the wireless technologies such as WiFi, RFID, cellular, etc. The listed papers in the last column of the table can leverage different wireless technologies as shown in Section~\ref{sub:sec:wireless:tech} for the attack. For example, one paper uses WiFi technology for attack and another paper uses cellular technology for attack. The stealthiness indicates if the attack is easy to detect. For example, in comparison to the active attacker who actively transmits wireless signals for attack, the passive attacker who passively receives the wireless signals is more stealthy. Even though passive attackers are considered to be stealthier, there are always detection approaches that can disable the stealthiness property of these passive attacks as shown in Section~\ref{sec:defenses}. Since these academic papers from the wireless sensing domain try to push the limit of sensing accuracy, they evaluate the attack performance from the perspective of sensing accuracy or localization error. The sensing accuracy reported in the state-of-the-art techniques for human activity or gesture recognition is usually more than 0.95~\cite{zhang2019towards, wang2015understanding} and the localization error is at the decimeter level~\cite{vasisht2016decimeter, ma2017minding}. The attack time in the wireless sensing attack can be defined as the time spent from deploying the attacking devices to successfully steal private information, which is not reported in these academic papers. We see some papers reporting the computational complexity of the sensing algorithms~\cite{hassanieh2014ghz}, which are not the attack time measured in real-world settings.

\subsection{Received Signals-based Inference Attacks}

The received wireless signals at the attacker can be used for inference attacks. Specifically, the attacker can collect the received wireless signals and then use them as features for an inference attack. For example, Zhu et al.~\cite{zhu2020tu} measure the variation of signal strength with a passive radio outside of the house to predict if there are occupants at home. IRshield~\cite{staat2022irshield} proposes to use the smart surface to distort the signal strength such that the attacker cannot predict the variation of the signal strength for occupant detection. However, the signal strength measurements are suffering from background noise. Vital-Radio~\cite{adib2015smart} and Wistress~\cite{ha2021wistress} (i.e., stress sensing) use the variation of the signal phase caused by the chest movement to achieve the inference attack, as the phase information is resilient to the noise but sensitive to the signal's traveling path.

\subsection{Wireless Channel-based Inference Attacks}

After obtaining the reconstructed wireless channel, the attacker can use it as the feature for an inference attack. Furthermore, the attacker can extract the features based on the reconstructed wireless channel for an inference attack. Specifically, the attacker can extract the following features based on the reconstructed wireless channel:
\begin{itemize}[leftmargin=*,itemsep=3pt]
\item 
\textbf{Wireless channel.}
The straightforward idea is to use the reconstructed features directly. Using the wireless channel as the features have been extensively studied to achieve gesture/activity recognition \cite{wang2016wifall, wang2017device, wang2016rt, palipana2018falldefi, virmani2017position, li2016csi} and indoor localization or tracking~\cite{xi2014electronic, feng2021rf, ali2015keystroke, chen2015tracking}. 
\item
\textbf{Signal attenuation.}
The signal attenuation can be directly derived from the signal's amplitude, which can characterize the wireless signal's power loss due to the over-the-air propagation. The signal attenuation feature has been widely used to infer human gestures/activities~\cite{cohn2012humantenna, srinivasan2008protecting, sigg2013rf, kosba2012rasid, sigg2013leveraging, wilson2010see, sigg2014telepathic, abdelnasser2015wigest,li2015human}, respiration/heart rate~\cite{ kaltiokallio2014non, patwari2013breathfinding, abdelnasser2015ubibreathe}, and localization/tracking~\cite{zhu2020tu, yu2018passive, seifeldin2012nuzzer, youssef2007challenges, moussa2009smart, xu2013scpl, lam2018new, lu2018simultaneous, bahl2000radar, paul2009rssi, wang2012robust, barsocchi2009novel}.

\item \textbf{Doppler shift.}
Doppler shift is caused by the victim's movements in the physical environment, which can be used as a feature to infer private information. A victim moving at a speed of $v$ at an angle of $\beta$ from the attacker in the physical environment experiences a Doppler frequency shift given by:
    \begin{equation}
     \Delta f \propto \frac{2v \cdot \cos \beta}{c} \cdot f_c.
    \end{equation}
The attacker can obtain the Doppler shift feature from the frequency-domain signals by applying the Fourier transform on the received signals. Prior works mainly leverage the Doppler shift for activity/gesture recognition and respiration/heart rate estimation using RF signals~\cite{ram2008doppler, ram2008through, lyonnet2010human, li2016non, li2008random, wang2011single, zhao2017noncontact, gu2009instrument, tan2016awireless, chetty2011through}.

\item \textbf{Time of Flight (ToF).}
 ToF, denoted by $\tau$, denotes the time duration during which the wireless signal travels through the physical environment for distance $d$, and is given by:
    \begin{equation}
        \tau=d/c.
    \end{equation}
    The estimation accuracy of the ToF information highly depends on the signal bandwidth $B$:
   \begin{equation}
        ToF \propto 1/B.
   \end{equation}
    In radar-based wireless sensing systems, ToF can be derived from the multipath profile describing the signal over time in a round trip. To conduct the inference attack, the attacker can snoop the pulse or frequency-modulated continuous-wave (FMCW) signals transmitted from the radar and reflected by the victim to create a multipath profile, which can be leveraged to infer the private information of hand gestures and location~\cite{adib20143d, adib2015multi, adib2015capturing, molchanov2015short, lien2016soli}. ML models have been employed in radar-based wireless sensing systems to analyze the collected 3D point clouds, which can achieve fined-grained sensing on emotion/gestures/activity/behavior recognition~\cite{zhao2016emotion, li2019making, vahia2020radio, fan2020home}, gait velocity and strait length estimation~\cite{hsu2017extracting}, sleep sensing~\cite{zhao2017learning, hsu2017zero, yue2020bodycompass}, human pose/mesh estimation~\cite{zhao2018through, zhao2019through}, 3D body skeleton~\cite{zhao2018rf}, human identification/authentication~\cite{vasisht2018duet, hsu2019enabling, fan2020learning, korany2019xmodal}, and respiration/heart rate detection~\cite{yue2018extracting}.

\item \textbf{Angle of Arrival (AoA) and Angle of Departure (AoD).}
AoA needs to be derived from the antenna array-enabled attacker. AoA of $l$-th signal path, denoted by $\theta_l(t)$, can be derived from the following equation:
    \begin{equation}
        d_e=D \cdot \cos\theta_l(t),
    \end{equation}
    where $d_e$ denotes the extra distance the signal travels, and $D$ denotes the antenna separation in the
    antenna array. Similarly, AoD can be derived at the Tx's antenna array. AoA and/or AoD information has been widely employed to achieve activity recognition and localization/tracking~\cite{wang2014rf, wang2013dude, adib2013see, melgarejo2014leveraging, zhang2010tasa, tanbo2015active, yu2020mmwave, nannuru2012radio, xie2019md, pu2013whole}.

\end{itemize}

\subsection{Discussion of Existing Inference Attacks}
\label{subsec:dis:xyz}

\myparatight{Discussion of received signal-based inference attacks} To use received signals for the attack, the attacker can simply infer human private information based on the machine learning-based network traffic pattern analysis. However, this received signal-based network traffic analysis suffers from the artifacts of communication data incorporated in the received signals. Usually, received signal-based wireless sensing attacks mainly leverage the signal strength of the received signals to infer the human private information, which suffers from the multipath effect in the indoor environment resulting in low sensing accuracy. Therefore, the received signal-based sensing attack usually focuses on the line-of-sight scenario, where the line-of-sight signals are dominant over the received signals in the with-device setting (e.g., a human holds a smartphone communicating with the WiFi access point). The multipath signal reflected off the human body should be resolved and used for inference attacks in device-free settings, where the victim does not co-locate with the transmitter.

\myparatight{Discussion of wireless channel-based inference attacks} Since the received signals-based sensing attacks are usually distorted by the communication data information, we would like to use the wireless channel to infer the human private information introduced by human movements. To use the estimated wireless channel for the attack, the attacker needs to accurately estimate the wireless channel. The attacker can simply infer the human private information based on the estimated wireless channel with machine learning models. This usually requires well-trained machine learning models on large-scale datasets, as the estimated wireless channel may not only be affected by human movements~\cite{ayyalasomayajula2020deep}. To this end, the signal path that is affected by the subject of interest should be extracted for attacking purposes, which requires the attacker to resolve the multipath over frequency, time, or space dimension. To do so, ToF can be leveraged to achieve high sensing accuracy by resolving the multipath in the frequency domain with a large bandwidth, and AoA/AoD can be leveraged to achieve sensing accuracy by resolving the multipath in the space domain with an antenna array. However, the attacker needs to be instrumented with a large antenna array or occupy a large frequency band, which will further burden the existing wireless spectrum usage.  An attacker can use signal attenuation derived from the estimated wireless channel for the attack, which is straightforward. However, it suffers from the multipath effect resulting in inaccurate attenuation estimation. Doppler shift is another factor that can be leveraged for sensing attack, while it is related to the moving speed of the subject of the target. As a result, Doppler shift cannot achieve fine-grained sensing attacks, even though the speed of the human movement is slow in practice.

\begin{figure}
  \centering
  \includegraphics[width=\linewidth]{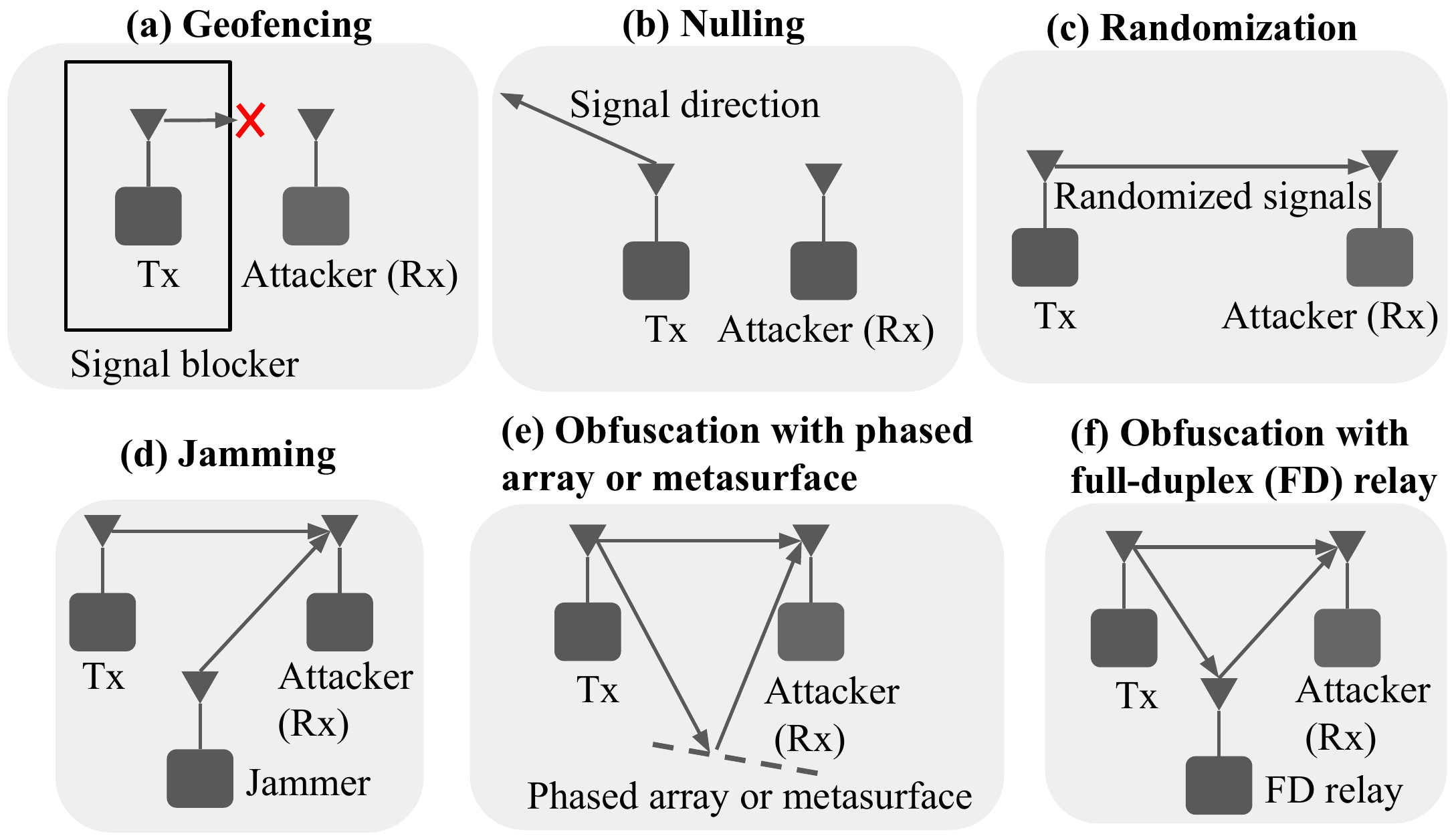} 
    \caption{Illustration of the prevention strategy. (a) geofencing that can block the wireless signals at the transmitter. (b) Nulling can nullify the signals received by the attacker. (c) Randomization introduces artifacts to the transmitted wireless signals. (d) Jamming can distort the received signals at the attacker. Obfuscation with a phased array or meta surface (e) and full-duplex relay (f) can distort the received wireless signals at the attacker.}
    \label{fig:sys:defense:prevention}
\end{figure}

\section{Taxonomy of Existing Defenses}
\label{sec:defenses}

\subsection{Prevention Strategy}
Fig.~\ref{fig:sys:defense:prevention} summarizes and illustrates the prevention strategies against the inference attacks in HCWS. Table~\ref{table:prevention:detection} (a) presents the taxonomy of prevention strategies against inference attacks. 

\subsubsection{Shielding Wireless Signals}

The root cause of the inference attack is due to the widespread propagation nature of wireless signals and the multipath effect in the physical environment, thereby any attacker residing in the coverage area of the Tx can sniff the wireless signals. To prevent the inference attack, we can shield the transmitted signals such that the attacker's Rx cannot receive them using the following two methods:
\begin{itemize}[leftmargin=*,itemsep=3pt]
\item
\myparatight{Geofencing} Geofencing is a way that can block the wireless signal so that it becomes inaccessible to the attacker. To do so, we can cover the walls with electromagnetic shielding paints, customize the wireless signal coverage with 3D fabricated reflectors~\cite{xiong2017customizing, cho2023mmwall, chen2021pushing} or backscatter arrays~\cite{li2019towards, zelaya2021lava}, as shown in Fig.~\ref{fig:sys:defense:prevention}(a).
\item
\myparatight{Nulling} To eliminate or mitigate the wireless signal propagation that is accessible to the attacker, the TX can also beamform the signal towards the desired Rx~\cite{dai2013eavesdropping} to minimize the signals leaking in the direction that could be received by the attacker, as shown in Fig.~\ref{fig:sys:defense:prevention}(b). Furthermore, if the location of the attacker is known, the Tx can apply beamforming to generate a deep null towards the attacker. Abedi et al.~\cite{abedi2021can} leverage the nulling capability of WiFi access points, and PushID~\cite{wang2019pushing} uses the blind beamforming to extend the coverage of the backscatter communication, which can be used to eliminate the eavesdropping in WiFi backscatter sensing systems.
\end{itemize}

\subsubsection{Obfuscating Wireless Signals}

To prevent inference attacks, we can also obfuscate the transmitted signals, such that the attacker cannot extract useful features from the sniffed wireless signals. To do so, the Tx can either randomize the transmitted signals or jam the received signals at the attacker's Rx as follows. 
\begin{itemize}[leftmargin=*,itemsep=3pt]
\item
\myparatight{Randomizing the transmitted signals} To obfuscate the transmitted signals, one way is to randomize the transmitted signals such that the attacker cannot predict anything from the traffic analysis based on the received signals as shown in Fig.~\ref{fig:sys:defense:prevention}(c). For example, RF-Cloak~\cite{hassanieh2015securing} randomizes the illuminated signals transmitted from the RFID reader to disable the attacker. Wijewardena et al.~\cite{wijewardena2020plug} consider randomization of the signal strengths to disable the attacker.
\item
\myparatight{Jamming the signals received by the attacker} Another way to obfuscate the transmitted signals is to deploy a signal generator to jam the received signals at the attacker, such that the signal-to-interference plus noise ratio (SINR) at the attacker is under the noise floor to disable the attacker, as shown in Fig.~\ref{fig:sys:defense:prevention}(d). For example, Jiao et al.~\cite{jiao2021openwifi} consider injecting artificial channels at the Tx to prevent inference attacks. Huang et al.~\cite{huang2020intelligent} use programmable metasurface to jam the pilot of the signals, and Lyu et al.\cite{lyu2020irs} use the programmable metasurface to jam the over-the-air signals.  
\end{itemize}

\begin{figure}
  \centering
\includegraphics[width=\linewidth]{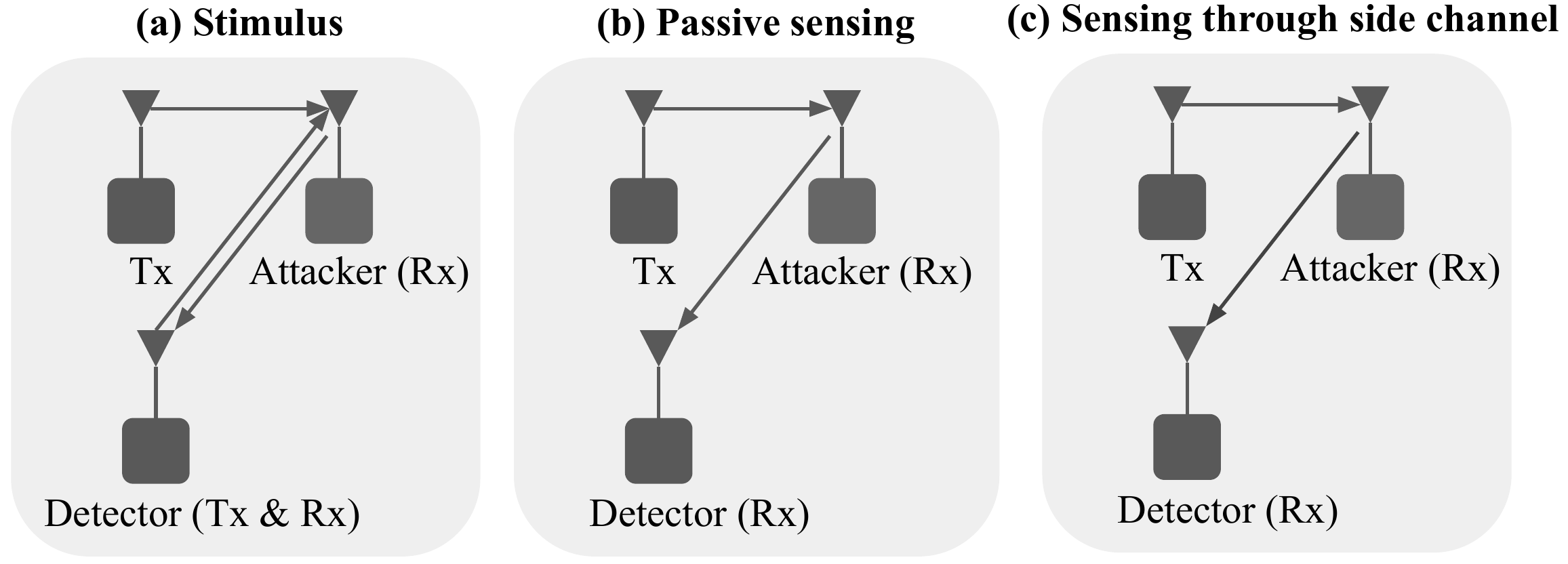} 
    \caption{Illustration of the detection strategy. (a) Stimulus uses the generated wireless signals to excite the attacker for detection purposes. (b) Passive sensing can detect the existence of the attacker by overhearing the emanations from him/her. (c) Sensing through the side channel can detect the attacker by sensing the leakage of the undesired side-channel information from the attacker's Rx.}
    \label{fig:sys:defense:detection}
\end{figure}

\subsubsection{Obfuscating the Wireless Channel}

Prevention methods mentioned above mainly focus on Tx-side shielding and obfuscation. The wireless channel plays an important role in human-centered wireless sensing and can also be obfuscated using techniques such as programmable phased arrays,  metasurfaces, or full-duplex relays. Obfuscating the wireless channel eventually leads to noisier wireless signals received by the attacker. 
\begin{itemize}[leftmargin=*,itemsep=3pt]
\item 
\myparatight{Reconfigurable phased array or metasurface-based wireless channel obfuscation} To obfuscate the wireless channel, we can use a reconfigurable phased array consisting of multiple discrete phase shifters that can change the phase of the wireless signals, as shown in Fig.~\ref{fig:sys:defense:prevention}(e). For example, LAIA~\cite{li2019towards} uses a phased array to control the wireless channel in the desired way by changing the wireless signal's phase.  We can also use the programmable metasurface to change the impinged signal's phase in the desired way. As such, the signals received by the attacker cannot help to extract the clean wireless channel that is only affected by the victim for private information inference. For example, IRShield~\cite{staat2022irshield} designs a metasurface that can change the wireless channel to disable eavesdropping. Hu et al.~\cite{hu2021ris} use the reconfigurable metasurface to change the wireless channel coefficients. Staat et al.~\cite{staat2021mirror} use the metasurface to achieve the jamming purpose that could disable eavesdroppers. 

\item
\myparatight{Full-duplex relay-based wireless channel obfuscation} Another way to obfuscate the wireless channel is to use full-duplex relays, as shown in Fig.~\ref{fig:sys:defense:prevention}(g). An amplify-and-forward (AF) relay amplifies and delays the impinging signal from the Tx and then forwards it to the attacker, during which the AF relay can change the amplitude and/or phase of the Tx signal. As such, the AF relay can change the wireless channel in the desired way such that the attacker cannot extract the desired and clean wireless signals affected by the victim for private information inference. For example, PhyCloak~\cite{qiao2016phycloak} uses the AF relay node to change the wireless channel that can prevent the attacker. Channel Spoofer~\cite{qiao2017channel} further demonstrates the AF relay node can change the wireless channel as designed. Sun et al.~\cite{sun2020destructive, sun2021destructive} use the AF relay to achieve destructive signal addition at the attacker in RFID-based sensing systems.
\end{itemize}

When the attacker is performing the active inference attack, the feasible defenses are jamming and obfuscation techniques. This is because the active attacker does not rely on the legitimate transmitter's transmissions to infer the human private information.

\begin{table*}
\Huge
\begin{center}
\subfloat[Prevention]{\scalebox{0.45}{
\begin{tabular}{|c|c|c|c|c|c|c|c|c|c|c|}
\hline
  \multicolumn{2}{|c|}{\textbf{Shielding}}  &
  \multicolumn{2}{|c|}{\textbf{Obfuscating wireless signals}} &
  \multicolumn{3}{|c|}{\textbf{Obfuscating wireless channel}} &
   \multicolumn{3}{|c|}{\textbf{Property}} &
  \multicolumn{1}{|c|}{\textbf{Papers}} \\
\hline
\hline
  Geofencing & Nulling & Randomization & Jamming & Phased array & Metasurface & FD relay & Cost& Stealthiness& Wireless tech.&\\
\hline
  $\cmark$ & $\xmark$  & $\xmark$ & $\xmark$ & $\xmark$ & $\xmark$ &  $\xmark$ & $\uparrow$&$\downarrow$ &- & ~\cite{xiong2017customizing, cho2023mmwall, chen2021pushing,li2019towards, zelaya2021lava} \\
  \hline
  $\xmark$ & $\xmark$  & $\xmark$ & $\xmark$ & $\cmark$ & $\xmark$ &  $\xmark$ & - & $\downarrow$ &- &~\cite{li2019towards,staat2021mirror,zelaya2021lava} \\
  
\hline
 $\xmark$ & $\cmark$ & $\xmark$ & $\xmark$ & $\xmark$  & $\xmark$ & $\xmark$ &- & $\downarrow$&- & ~\cite{dai2013eavesdropping,yeh2020security,abedi2021can,wang2019pushing}\\
\hline 
 $\xmark$ & $\xmark$ & $\cmark$ & $\xmark$ & $\xmark$ & $\xmark$ & $\xmark$ &- &$\downarrow$&-&~\cite{hassanieh2015securing, wijewardena2020plug} \\
\hline
 $\xmark$ & $\xmark$ & $\xmark$ & $\cmark$& $\xmark$ & $\xmark$ & $\xmark$ & -&$\downarrow$&-&~\cite{staat2021mirror,jiao2021openwifi, shenoy2022rf,wijewardena2020plug,huang2020intelligent,lyu2020irs} \\
\hline
 $\xmark$ & $\xmark$ & $\xmark$ & $\xmark$& $\xmark$ & $\cmark$ & $\xmark$ &$\uparrow$ & $\uparrow$&-&~\cite{cho2023mmwall, staat2022irshield, hu2021ris, huang2020intelligent, staat2021mirror} \\
\hline
 $\xmark$& $\xmark$ & $\xmark$ & $\xmark$ & $\xmark$ & $\xmark$ & $\cmark$ & $\uparrow$ &$\uparrow$& -&
~\cite{qiao2016phycloak,qiao2017channel,sun2021destructive, sun2020destructive} \\

\hline

\end{tabular}
}}

\subfloat[Detection]{\scalebox{0.45}{
\begin{tabular}{|c|c|c|c|c|c|c|}
 \hline
  \textbf{Stimulus} & \textbf{Passive sensing} & \textbf{Side-channel}  &
  \textbf{Cost} & 
  \textbf{Stealthiness} &
  \textbf{Wireless tech.} &
  \textbf{Papers}  \\
\hline
\hline
$\xmark$ & $\cmark$ & $\xmark$ & - & $\downarrow$ & - &~\cite{chaman2018ghostbuster,manco2015detection,mukherjee2012detecting,park2006hidden,park2010rf,sharma2022lumos,srinivasan2008protecting,wild2005detecting,cheng2018dewicam,yeh2020security, ghasempour2020leakytrack} \\ 
\hline                                  

 \hline
$\cmark$ & $\xmark$ & $\xmark$ & - & $\uparrow$&-&~\cite{liu2018detecting,wang2014eyes,cheng2018dewicam,stagner2011practical, stagner2012locating,camera_detector,camera_detector_1, seguin2009detection, stagner2013detecting, thotla2013detection, li2018eye}  \\

 \hline 

 \hline
 $\xmark$ & $\xmark$ & $\cmark$ & - & $\uparrow$  & - &~\cite{cui2020sniffing,zhao2017noncontact} \\
 \hline     
\hline
\end{tabular}
}}
\caption{Taxonomy of existing prevention (a) and detection (b) strategies for the defenses against the inference attacks. $\cmark$: used, $\xmark$: not used, -:all possible cases, $\uparrow$: high, and $\downarrow$: low.}
\label{table:prevention:detection}
\vspace{-1.0\baselineskip}
\end{center}
\end{table*}

\subsection{Detection Strategy}
Detection of inference attacks aims to detect an attacker's Rx, which is challenging because the passive inference attack only passively sniffs the wireless signals in the environment without transmitting any signals. Detecting an attacker's Rx can be viewed as a sensing problem, where the detector aims to sense the Rx used and deployed by the attacker. To this end, there are three methods for detecting an Rx (\emph{i.e.}, attacker), as illustrated in Fig.~\ref{fig:sys:defense:detection}. We present the taxonomy of detection strategies against the inference attacks in Table~\ref{table:prevention:detection}(b).

\begin{itemize}[leftmargin=*,itemsep=3pt]
\item
\myparatight{Stimulus} Although the attacker's passive Rx does not actively emit any signal, we can actively transmit a known stimulation signal that can trigger the attacker's Rx circuit to leak unintended signals, which can then be captured for detection purposes, as shown in Fig.~\ref{fig:sys:defense:detection}(a). For example, many research papers~\cite{seguin2009detection, stagner2013detecting, stagner2011practical, stagner2012locating, thotla2013detection, li2018eye} show that by actively transmitting a known stimulation signal, the attacker's circuit can be triggered to reflect the unintended wireless signals, which could be further analyzed to detect the attacker. Recent works~\cite{camera_detector, camera_detector_1, liu2018detecting} also show that by emitting light signals, hidden cameras can be detected.

\item 
\myparatight{Passive sensing}
The passive devices deployed by the attacker can still leak the wireless signals, although it is inactive and just listening. So, we can sense these weak signal leakage from the attacker to detect the presence of the inference attack as shown in Fig.~\ref{fig:sys:defense:detection}(b). For example, many research papers~\cite{mukherjee2012detecting, park2010rf, wild2005detecting, park2006hidden, manco2015detection,sharma2022lumos, chaman2018ghostbuster, cheng2018dewicam} demonstrate and analyze the signal leakage from the local oscillator of the radio that can be sensed to detect the attacker. Recent works~\cite{yeh2020security, ghasempour2020leakytrack} show the security issue of the leaky wave antennas in Terahertz communication and sensing, which can be detected to eliminate the attack. 

\item 
\myparatight{Sensing through side-channel}
A passive device that does not actively transmit any signal can also leak the signals through side channels. Therefore, we can detect the presence of the attacker over these side channels, as shown in Fig.~\ref{fig:sys:defense:detection}(c). For example, Cui et al.~\cite{cui2020sniffing} use a wireless signal sniffer to detect the signal leakage of the visible light communication and sensing systems. 
\end{itemize}

Since the active attacker needs to transmit the wireless signals and analyze the backscattered signals for inference attack, it is easy to detect them through passive sensing and sensing through side channels.

\section{Challenges for Privacy-preserving HCWS}
\label{sec:challenges}

\noindent \textbf{C1: Sniffing device deployment.} The active attackers can always transmit known wireless signals to infer human private information, while the passive attackers need to rely on the existing wireless signals transmitted by the deployed wireless sensing systems for inference attacks. However, passive attacks are more covert than active attacks. As a result, passive attacks are difficult to detect. Active attacks are easy to detect and localized by analyzing the transmitted wireless signals from the active attackers.

Note that the existing passive attacks usually assume the attacker knows the exact signal type and frequency band the wireless sensing systems have used, which is not realistic for deploying real-world inference attacks. From the defense perspective, the signal type and frequency band used by the wireless sensing systems are also private information. If we can protect this information from being leaked, we can fundamentally defend against passive attacks.

\noindent \textbf{C2: Compensating hardware imperfection and artifacts.} The hardware imperfection of the transceiver introduces an extra phase shift $\phi(t)$, and the moving transceiver or reflectors will introduce phase shift $\gamma(t)$ due to the Doppler shift effect. All these changes are collectively referred to as the wireless channel. Therefore, for the signal transmitted at a carrier frequency of $f_c$ (or with wavelength $\lambda=\frac{c}{f_c}$ where $c$ is the speed of light), the single-path wireless channel $h(t)$ can be defined as:
\begin{equation}
    h(t)=\alpha(t) \cdot \exp \left(-j 2\pi \frac{d(t)}{\lambda}+j\phi(t)+j\gamma(t) \right).
\end{equation}
In a real-world wireless environment, the signal received at the Rx is a composition of multiple copies of the original signal due to the multipath effect, where each copy can experience different attenuation, delay, and/or phase change. We can represent the channel seen by the Rx as the combination of all the possible $L$ single-path channels:
\begin{equation}
\label{eq:multi:channel}
    h(t)=\sum_{l=1}^{L} \alpha_{l}(t) \cdot \exp \left( -j 2\pi \frac{d_l(t)}{\lambda}+j\phi(t)+j\gamma_l(t) \right).
\end{equation}
We are only interested in $d_l(t)$ or $\gamma_l(t)$, which is related to the subject of interest. As a result, it is highly challenging to resolve the composited signals received at the receiver due to the multipath effect. The hardware imperfection introduced by the transmitter is hard to compensate for, as the attacker cannot obtain the transmitter's hardware artifacts. As this hardware imperfection is unique to the hardware itself, it's usually leveraged for hardware fingerprinting.

The attacker needs to eliminate the human-introduced artifacts that are hidden in the wireless signals. For example, different people could perform the same activity or gesture with different scales and/or orientations with respect to the attacker. To remove the human-introduced artifacts in the extracted features, the attacker can rescale the time-series features~\cite{zhang2016mudra, mcglynn2010ensemble, muller2007dynamic}. To remove the orientation artifacts in the extracted features, the prior works mainly leverage the space diversity by using two antennas to receive the wireless signals based on the fact that the orientation artifact can be canceled out across different antennas~\cite{sun2021orientation, zhang2019towards}. After the pre-processing, the attacker can use them as the input of private information inference components for indoor localization~\cite{yiu2017wireless, jain2021low} and tracking~\cite{aravinda2021optimization}.

\noindent \textbf{C3: Vulnerable machine learning-based private information inference.} Recently, we find that deep learning has been extensively studied in human-centered wireless sensing for high sensing accuracy without considering privacy leakage. Therefore, it is important to build trustworthy deep-learning models and apply them to the existing signal-processing pipeline of human-centered wireless sensing. We identify the following gaps or challenges to achieve privacy-preserving ML-enabled human-centered wireless sensing systems. Under our signal processing pipeline-based framework, we find that the wireless sensing systems often leverage machine learning models for human private information inference, which are vulnerable to adversarial attacks~\cite{szegedy2013intriguing, carlini2017towards}. Specifically, the attacker can add small carefully crafted noises to wireless signals to turn them into adversarial examples, which can obfuscate the machine learning models employed by the legitimate transceiver, such that the legitimate wireless sensing systems would make random inferences about human's private information. Even though we can directly apply the existing defensive mechanisms from the trustworthy machine learning community to secure the machine learning models used in HCWS, it is challenging to integrate these defensive mechanisms from the end-to-end HCWS system design point of view. This is because the existing defensive mechanisms for machine learning models are only designed for machine learning models without considering the integration and role of these models in an end-to-end system.

\noindent \textbf{C5: Resolving multipath in a dynamic and multiple person environment.} The prior works on human-centered wireless sensing mainly focus on one subject of interest in a quasi-static wireless environment. This is because wireless sensing mainly leverages the variation of the wireless environment affected by human movements to infer human private information. When there are multiple different reflectors (e.g., walls, chairs, furniture, etc.) or moving artifacts in the environment, the received wireless signals at the attacker will be distorted. So, it is important for the attacker to resolve the multipath and extract the signal path that is only affected by the subject of interest. We illustrate the pros and cons of the following multipath resolving approaches from the time, frequency, and space domains.  
\begin{itemize}[leftmargin=*,itemsep=3pt]
\item
\myparatight{Resolving multipath in the time domain}
To eliminate the artifacts introduced by the wireless environment, the straightforward idea is to assume the wireless environment is only affected by the victim and all other objects are relatively static. Specifically, the signal cancellation approach can be employed to cancel out the effects from all the other non-victims (\emph{e.g.}, walls, desks), while this approach barely works in the dynamic environment. This is because we cannot assume the environment-introduced artifacts are not changing over time for our cancellation purpose.  
\item
\myparatight{Resolving multipath in the frequency domain}
The main idea of resolving multipath in the frequency domain is to leverage the characteristic of the frequency-selective wireless channel in which the wireless signals operating at different frequencies will be affected by the physical environment differently. To do so, we leverage wireless signals that occupy a wide frequency band to measure the time-of-flight for resolving the multipath, while the wide-band signals are barely available due to the limited wireless spectrum. 
\item 
\myparatight{Resolving multipath in the space domain}
Resolving the multipath in the space domain is intuitive, as the different objects in the physical environment will be located in different places. Therefore, the signals reflected by these different objects will undergo different physical paths, resulting in different AoA values that can be measured to resolve the multipath signal propagation. However, using multiple arrays will introduce deployment costs. Given the receiving antenna array, the AoA resolution is limited by this array's aperture size. When two objects are close to each other, they will introduce similar signal propagation paths that cannot be resolved over the space domain.  
\end{itemize}

\noindent \textbf{C6: Obfuscating the attacker without affecting the legitimate receiver's sensing purposes.} Wireless sensing can be leveraged for human-computer interaction, smart homes, and asset tracking. So, it is important to obfuscate the attacker without affecting the legitimate receiver's sensing purposes. However, this is very difficult and challenging, as the attacker and legitimate receiver share the same wireless environment. As a result, the legitimate receiver will also receive these distorted wireless signals. The distorted signal cancellation at the legitimate receiver will also introduce extra artifacts that are hard to eliminate. Since we do not know where the attacker is, it is not possible to shine the very narrow beam toward the attacker without affecting the legitimate receiver's sensing purposes.

Shielding wireless signals and wireless channels will affect and even suspend normal wireless communication due to the weak received signal strength at the legitimate receiver, which will not be desirable for joint communication and sensing systems as wireless sensing is usually a byproduct of wireless communication. In comparison to the prevention methods, the detection methods (e.g., passive sensing and sensing through side channels) will not affect the existing wireless communication, while it is very difficult to detect the leaked side-channel information from the attacker that is weak and usually under noise floor.

\section{Discussions and Future Directions}
\label{sec:discussion}

\begin{figure}
  \centering
  \includegraphics[width=\linewidth]{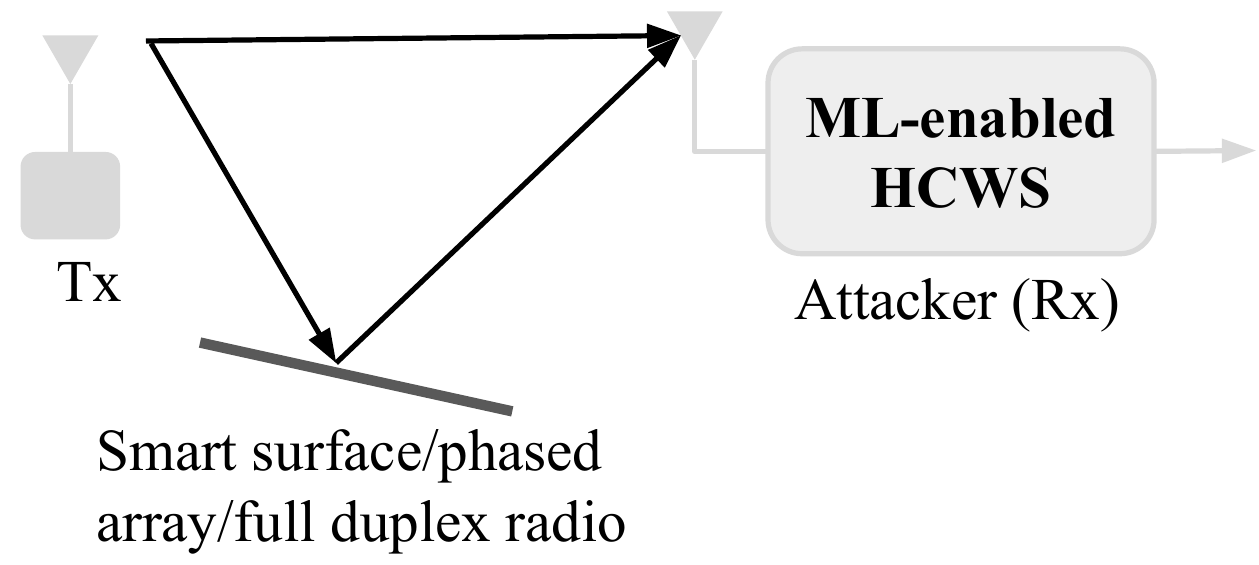} 
    \caption{Adversarial example introduced by the smart surface, phased array, or full-duplex radio can disable the private information inference at the attacker.}
    \label{fig:pets:adv}
\end{figure}

\noindent \textbf{Applying trustworthy machine learning.} The existing trustworthy machine learning models do not take into account privacy issues when they are integrated into human-centered wireless sensing systems~\cite{ayyalasomayajula2020deep, xu2019faho}. For example, we know that adversarial examples can be leveraged to obfuscate the machine-learning models, while we still do not know how to apply the adversarial examples to achieve privacy-preserving ML-enabled human-centered wireless sensing systems from the end-to-end system design point of view. One possible solution is to generate the adversarial examples at the input features of the attacker's machine learning models without considering the end-to-end HCWS design, which requires us to access the attacker's sensing system. Since the input features of the machine learning models are coming from the signal processing pipeline, we can introduce over-the-air adversarial examples with smart surface or full-duplex relay nodes. 
\begin{itemize}[leftmargin=*,itemsep=3pt]
\item
\myparatight{Adversarial examples added to the wireless channel} To defend against an inference attack conducted based on the wireless channel, we can turn it into adversarial examples via deploying the programmable smart surface or full-duplex radio in the physical environment, such that the adversarially perturbed wireless channel makes the attacker's machine learning models randomly and incorrectly predict a victim's private information as shown in Fig.~\ref{fig:pets:adv}. The recent paper presents WiADv~\cite{zhou2022wiadv}, a system that uses the full-duplex radio to obfuscate the estimated wireless channel at the receiver of the wireless sensing-based gesture recognition systems. 
\item
\myparatight{Adversarial examples added to the received signals} To defend against an inference attack conducted based on the received signals, the adversarial examples could be generated by a generator (i.e., full-duplex node) to introduce misclassification to the attacker's machine learning models. In this case, the wireless signals received by the attacker's Rx consist of the signals transmitted from the legitimate Tx and the signals generated by the generator. In other words, the composition of the signals transmitted from the legitimate Tx and the signals generated by the generator should be adversarial examples to the attacker's machine learning models. Such defenses are illustrated in Fig.~\ref{fig:pets:adv}. For example, a recent paper proposes RF-Protect~\cite{shenoy2022rf}, a system that uses the smart surface to obfuscate the radar-based human activity recognition systems by generating ghost reflections.
\end{itemize}

To protect a legitimate Rx from being affected by these adversarial examples, it could use different mechanisms from the attacker to analyze the received wireless signals. In particular, a legitimate Rx may know the added adversarial perturbations and filter them before analyzing the wireless signals if the Rx and the Tx have established a secure communication channel in advance and can exchange the added adversarial perturbations. Moreover, the generated adversarial examples can be directed to the attacker's Rx without interfering with the legitimate Rx's sensing purpose using either directional antennas or beamforming techniques. In particular, we know the locations of the legitimate Rx and thus we can direct the adversarial examples towards directions not covering the legitimate Rx. Furthermore, it is an interesting future research direction to carefully design the adversarial examples, such that the legitimate Rx's analysis is unaffected by the adversarially perturbed wireless signals while the attacker's machine learning models make random and even incorrect inferences based on the adversarially perturbed wireless signals.

\noindent \myparatight{Defenses with formal privacy guarantees} Existing defenses do not have formal privacy guarantees. For instance, the prevention strategies with wireless signal obfuscation simply add noise to the signals received by the attacker without considering the privacy guarantee. When the attacker employs machine learning models for human private information inference, this added noise can be mitigated through adversarial training or incoherent averaging over multiple received signals. The detection strategies mainly focus on detecting the signal leakage at the attacker's Rx. Therefore, the defenses may be broken by advanced and adaptive inference attacks that know these defenses. Therefore, it is important to generate the noise derived from the differential privacy mechanisms~\cite{dwork2014algorithmic}, which can provide a privacy guarantee. Moreover, we could also leverage differential privacy and analyze the tradeoff between the privacy guarantee and the utility of wireless sensing or communication to achieve joint sensing and communication or defenses without affecting the legitimate transceiver's sensing purpose.

\noindent \myparatight{Multimodal sensor fusion-based inference attacks} Existing inference attacks only leverage wireless signals from a single Rx. To be resilient and robust to the dynamic and multipath wireless environment, the attacker can leverage multimodal sensor fusion, in which multiple Rxs can be used to sense the variation of the physical environment. As such, this multimodal sensor fusion provides improved diversity for the attacker to infer private information about the victim. To mitigate the privacy leakage in human-centered wireless sensing, we can still leverage the above defensive mechanisms. This is because multimodal sensor fusion highly depends on trustworthy signal sources from different devices. The above defensive mechanisms can also defend against the inference attack on each individual device in multimodal sensor fusion-based inference attacks. However, how effective using the above defensive mechanisms against the multimodal sensor fusion-based inference attacks needs further exploration. Moreover, One great challenge of multimodal sensor fusion-based inference attacks is data stream synchronization, as these multimodal features are extracted from multiple devices.

\noindent \myparatight{Detecting inference attack based on the estimated wireless channel}
We identify that existing detection methods haven't leveraged wireless channels. It is an interesting future research direction to explore wireless channel-based detection methods. For instance, we can detect an attacker's Rx by measuring the wireless channel wireless channel. One idea is that the existence of the attacker's Rx changes the multipath reflection profile of the wireless channel. This is because wireless signal propagation highly depends on the reflection of different objects in the physical environment. Therefore, by comparing the difference of the multipath profile of the physical environment, we can detect the attacker's Rx. However, this highly depends on the granularity of the multipath profile. We believe that advanced sensors (e.g., LiDAR or mmWave Radar) can be used to create the 3D point cloud of the environment and then leverage computer vision techniques to identify the attackers.

\section{Conclusions}
\label{sec:cons}
In this work, we systematized the literature on human-centered wireless sensing-based inference attacks and defenses through frameworks and insights. To do so, we propose a signal processing pipeline-based framework to bridge the gap between wireless sensing and privacy implications. Then, we instantiate the wireless sensing-based inference attacks and defenses. Based on this, we address the open challenges and identify the design space for privacy-preserving wireless sensing.

\begin{acks}
This research received no specific grant from any funding agency in the public, commercial, or not-for-profit sectors. We are grateful to the anonymous reviewers and our shepherd for their insightful feedback.
\end{acks}

\bibliographystyle{ACM-Reference-Format}
\bibliography{sample-base}

\end{document}